\documentclass{PoS}

\usepackage{amsmath}
\usepackage{amssymb}
\usepackage{amstext}
\usepackage{amsfonts}
\usepackage{subfigure}
\usepackage{multirow}
\usepackage{epsfig}
\usepackage{amssymb}
\usepackage{epsf}
\usepackage{epsfig}

\newcommand{\ba}{\begin{eqnarray}}
\newcommand{\ea}{\end{eqnarray}}
\newcommand{\be}{\begin{equation}}
\newcommand{\ee}{\end{equation}}
\newcommand{\bd}{\begin{displaymath}}
\newcommand{\ed}{\end{displaymath}}
\newcommand{\een}{\nonumber\end{equation}}
\newcommand{\bea}{\begin{eqnarray}}
\newcommand{\eean}{\nonumber\end{eqnarray}}
\newcommand{\eea}{\end{eqnarray}}

\def\eq#1{Eq.~(\ref{#1})}

\def\fig#1{Fig.~\ref{#1}}

\def\tab#1{Table~\ref{#1}}

\newcommand{\gap}{\hspace{10pt}}

\newcommand{\mev}{\mathrm{MeV}}

\newcommand{\fm}{~\mathrm{fm}}

\newcommand{\mps}{m_{\rm{PS}}}

\newcommand{\hbchipt}{\rm{HB}\chi\rm{PT}}

\newcommand{\beq}{\begin{equation}}   
\newcommand{\eeq}{\end{equation}}   
\newcommand{\beqn}{\begin{eqnarray}}  
\newcommand{\eeqn}{\end{eqnarray}}

\def\mcO{{\mathcal O}}

\def\la{\langle}
\def\ra{\rangle}

\newcommand{\old}[1]{}

\title{The quark contents of the nucleon and their implication for dark matter search}

\ShortTitle{Quark contents of the nucleon}

\author{C. Alexandrou$^{ab}$, M. Constantinou$\,^b$, \speaker{V. Drach}$\,^{cd}$, K. Hadjiyiannakou$^b$, K. Jansen$\,^c$ , G. Koutsou$^a$,~A. Strelchenko$\,^b$ and A.~Vaquero$^a$ \\
\llap{$^a$} Computation-based Science and Technology Research Center (CaSToRC), The Cyprus Institute, \\
20 Constantinou Kavafi Street Nicosia 2121, Cyprus \\
\llap{$^b$} Departament of Physics, University of Cyprus, P.O. Box 20537, 1678 Nicosia, Cyprus \\
\llap{$^c$}{NIC, DESY Zeuthen, Platanenallee 6, D-15738 Zeuthen, Germany\\}
\llap{$^d$}{$CP^3$-Origins \& the  Danish Institute for Advanced Study  DIAS, University of Southern Denmark, Campusvej 55, DK-5230 Odense M, Denmark\\}
E-mail:  \email{drach@cp3.dias.sdu.dk}
}

\abstract{We present results concerning the light and strange quark
  contents of the nucleon using $N_f=2+1+1$ flavours of maximally
  twisted mass fermions. The corresponding $\sigma$-terms are casting
  light on the origin of the nucleon mass and their values are
  important to interpret experimental data from direct dark matter
  searches. We discuss our strategy to estimate systematic
  uncertainties arising in our computations. Our preliminary results
  for the $\sigma-$terms read  $\sigma_{\pi N} = 37(2.6)(24.7)  \mev$ 
  and $\sigma_s=28(8)(10) \mev$.
We present our recent final analysis of  the $y_N$
parameter and found $y_N=0.135(46)$ including systematics\cite{Alexandrou:2013nda} .
}

\FullConference{31st International Symposium on Lattice Field Theory - LATTICE 2013\\
		July 29 - August 3, 2013\\
		Mainz, Germany}

\begin{document}

\section{Introduction}

The various evidences for the existence of dark matter have led to the development of 
experiments dedicated to detect dark matter directly. The detection relies on the 
measurements of the recoil of atoms hit by a dark matter candidate. One popular class of 
dark matter models involve an interaction between a WIMP and a nucleon mediated by a Higgs exchange.  
Therefore, the scalar quark 
contents of the nucleon are fundamental ingredients in the WIMP-nucleon 
cross section. In this way, the uncertainties of the scalar quark contents translate directly  
into the accuracy of the constraints on beyond the standard model physics. 
Since the coupling of the Higgs to quarks is, through the 
Yukawa interaction, proportional to the quark masses, 
it is important to know how large the scalar quark matrix elements of the nucleon are, 
in particular for the strange and charm quarks.  

One common way to write the parameters entering the relevant cross section are the 
so-called $\sigma$-terms of the nucleon:
\be\label{eq:sigma_terms}
\sigma_{\pi N} \equiv m \la N|  \bar{u} u+  \bar{d} d |N \ra \gap\textmd{and}\gap 
\sigma_s  \equiv  m_s\la N|  \bar{s}s |N \ra\; ,
\ee
where $m$ denotes the light quark mass and $m_s$ the strange quark mass. 
To quantify the scalar strangeness content of the nucleon 
a parameter $y_N$ is introduced,

\be\label{eq:y_para}
y_N \equiv  \frac{ 2\la N| \bar{s} s |N \ra }{ \la N| \bar{u} u+ \bar{d} d |N \ra},
\ee
which can be also related to the $\sigma$ terms of the nucleon in eq.~(\ref{eq:sigma_terms}).

The direct computation of the above matrix elements is known to be challenging on the lattice for 
several reasons. First, it involves the computation of "singlet"  or "disconnected'' diagrams 
that are very noisy. Second, discretisations that break chiral symmetry generally 
suffer from a mixing under renormalization between the light and strange sector, 
which is difficult to treat in a fully non-perturbative way.

However, as has been shown in \cite{Dinter:2012tt}, twisted mass fermions offer two 
advantages here: they provide both an efficient variance noise reduction for 
disconnected diagrams~\cite{Jansen:2008wv} and avoid the chirally violating 
contributions that are responsible for the mixing under
renormalization in our setup.
Note that a great effort has been spent developing techniques to
estimate efficiently the relevant disconnected contribution (see for instance\cite{Alexandrou:2012py,Engelhardt:2012gd,Oksuzian:2012rzb,Bali:2012qs,Alexandrou:2013wca}).

In this work we present a preliminary analysis of the systematic errors
occurring in the computation of
the $\sigma$-terms defined in \eq{eq:sigma_terms}. More precisely we
study the uncertainties associated to the excited states contamination,
lattice cut-off effects and the chiral extrapolation. We also summarise
our final results including all systematics for the $y_N$ parameter as
obtained in \cite{Alexandrou:2013nda}.



\section{Results }

\subsection{Lattice techniques}
In this study we use gauge configuration generated by the ETM
collaboration. We use $N_f =2+1+1$ ensembles with a number of light
quark masses corresponding to pseudoscalar meson masses ranging from
$220~\mev$ to $490~\mev$ and two lattice spacing, $a = 0.082 \fm$ and $a
= 0.064 \fm$,  to examine lattice cut-off effects.

We refer to \cite{Baron:2010bv} for details on the gauge ensemble used in this work. 
In order to compute matrix elements involving strange quarks, we work within a 
mixed action setup introducing an additional doublet of mass degenerate 
twisted mass quarks of mass $m_s$ in the valence sector.

The scalar quark matrix elements involved in \eq{eq:sigma_terms} can then 
be computed using the asymptotic 
behaviour of a suitable ratio of three and two-point functions defined as 
\be
\label{eq:ratio_def}
R_{O_q}(t_s, t_{\rm op}) = \frac{C^{O_q}_{\rm 3pts}(t_s,t_{\rm
    op})}{C_{\rm 2pts}(t_s) } = \la N | O_q | N \ra^{(\rm bare)} +
\mcO( e^{-\delta m  t_{\rm op}}) +  \mcO( e^{-\delta m (t_s - t_{\rm op})} )\;,
\ee
where $O_q$ refers to the operator in which we are interested in, 
namely $O_l \equiv \bar{u}u + \bar{d}d$ and $O_s\equiv\bar{s}s$. In
eq.~(\ref{eq:ratio_def}), $t_s$ refers to the source-sink 
separation and $t_{\rm op}$ to the source-operator separation. 
In addition, $\delta m$ stands for the mass gap between the nucleon and its first excited state.  
From eq.~(\ref{eq:ratio_def}) it is clear that large times $t_{\rm op}$ and $t_s$ are 
needed to suppress the so-called excited state contributions. 
However, due to the exponential decrease of the signal-over-noise ratio at large times, it is 
numerically very expensive to obtain a good signal for increasing $t_s$ or $t_{\rm op}$. 

The nucleon states themselves are created using smeared interpolating fields, which have 
already been optimized themselves to suppress excited state contaminations in the two point function. 

For the precise expression of the operators  $O_q$, their (solely) multiplicative renormalization 
properties and 
our computational techniques we refer the reader to
\cite{Dinter:2012tt}. In the following we discuss our analysis
strategy to obtain results extrapolated to the physical pion mass and
to estimate the systematic errors.

\subsection{$\sigma-$terms}

\begin{figure}[h]
\vspace*{-0.7cm}
\begin{minipage}[ht]{7cm}
\includegraphics[height=7.cm,width=7.cm]{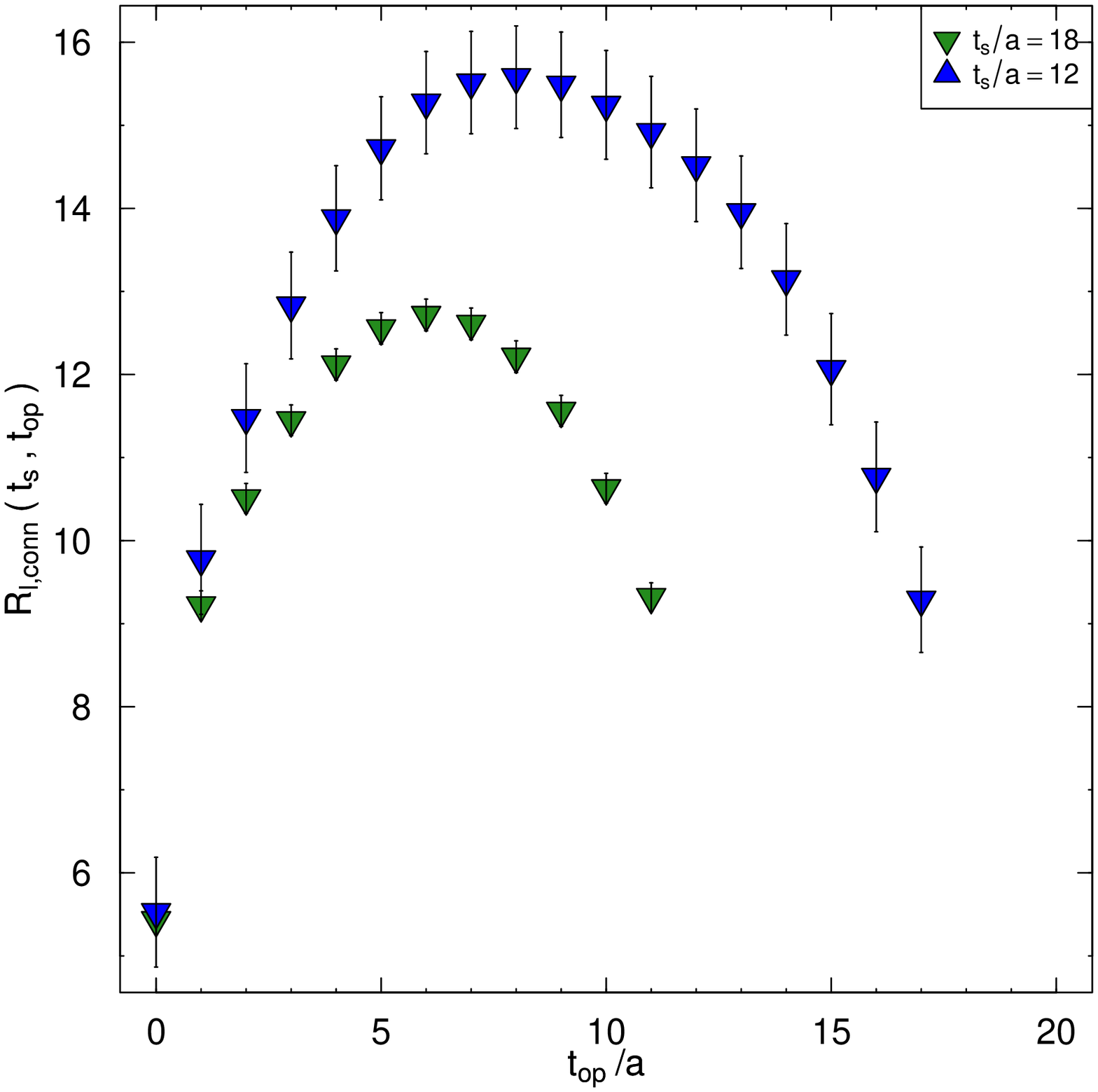}
\vspace*{-1.3cm}
\caption{$R_{O_l}$ for two different
  source-sink separations $t_s=12a \sim 0.98\fm$ and $t_s=18a\sim1.48 \fm$ on a $32^3\times64$
  lattice with a pseudoscalar mass of $\sim 355~\mev$}\label{fig:Rconn}
\end{minipage}
\hspace{0.5cm}
\begin{minipage}[ht]{7cm}
\includegraphics[height=7cm,width=7.cm]{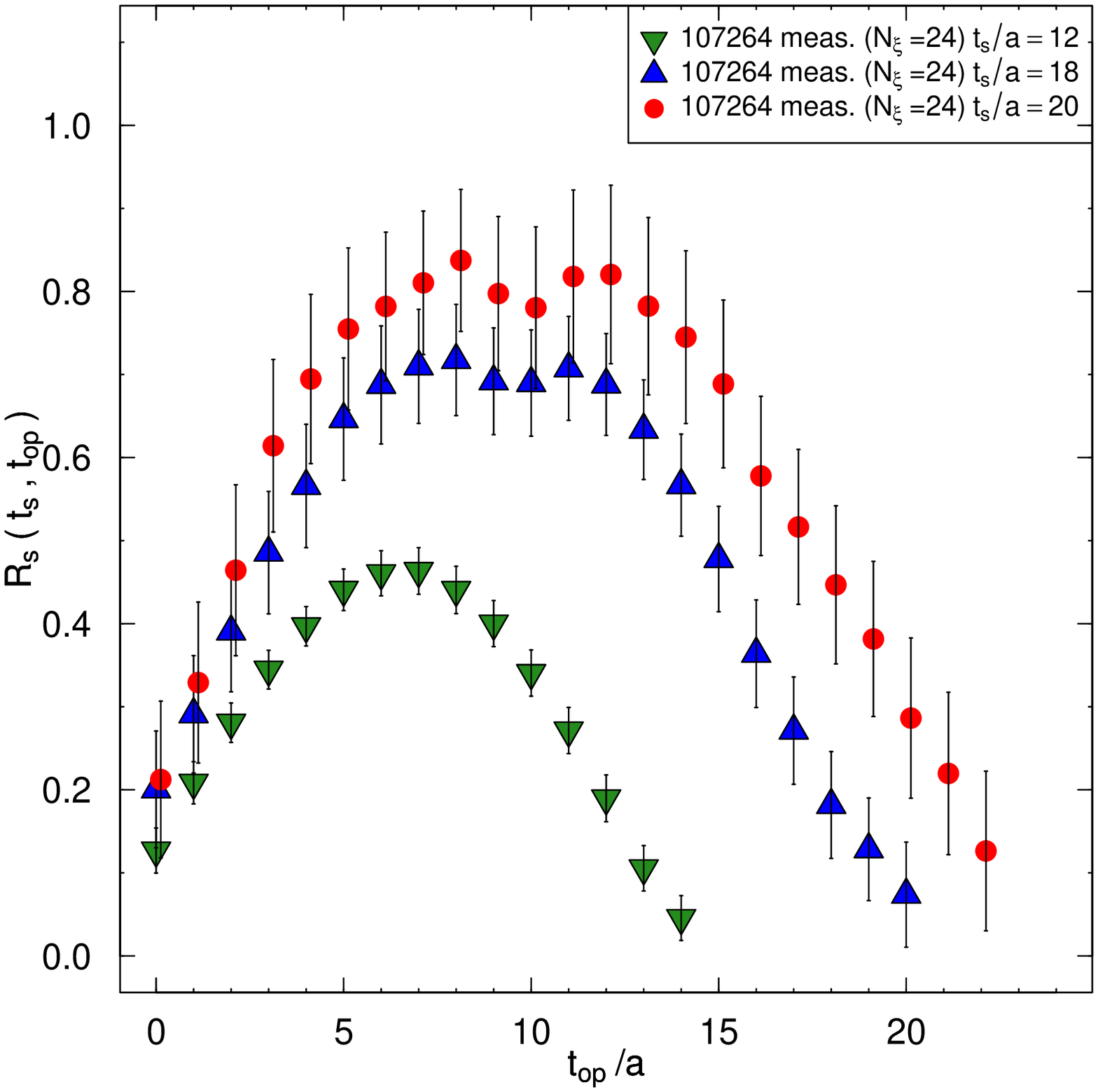}
\vspace*{-1.3cm}
\caption{$R_{O_s}$ for different
  source-sink separations ranging from $t_s\sim 0.98 \fm$ to $t_s \sim
  1.6 \fm$on a $32^3\times64$   lattice with a pseudoscalar mass of $\sim 392~\mev$}\label{fig:Rdisc}
\end{minipage}
\end{figure}

As discussed in \cite{Alexandrou:2012gz},
contributions from the excited states in \eq{eq:ratio_def} are not 
negligible both in $R_{O_l}$ and $R_{O_s}$ . We illustrate this in
\fig{fig:Rconn} and \ref{fig:Rdisc}, where we show the dependence of the ratio
\eq{eq:ratio_def} when $t_s$ is increased for $R_{O_l}$ (left) and $R_{O_s}$(right). In
the light sector we observe a shift of $\sim 22 \%$ in the value of
$R_{O_l}(t_s,t_s/2) $ when increasing $t_s$ from
$0.98 \fm$ to $1.48\fm$. In the strange sector, increasing $t_s$
from $0.98\fm$ to $1.6\fm$, changes $R_{O_s}(t_s,t_s/2)$ by about $\sim
80\%$.
Therefore, with the improved statistics employed here, it is not clear 
that the (time) asymptotic regime of \eq{eq:ratio_def} is reached.  
In order to nevertheless estimate the asymptotic values 
of the matrix elements, we performed extrapolations
of the lattice data to infinite source-sink separations. This step requires extrapolation that
leads to an additional systematic error previously neglected in other
computations.

We first describe our strategy to extrapolate the lattice data of
$R_{O_s}$ to $t_s\to\infty$ to infinite source-sink separation. 
Since we have data for every source-sink
separation $t_s$, we choose to perform fits to the following
function :
\be
f(t_s,t_{\rm{op}}) \equiv A(t_{\rm{op}}) + B(t_{\rm{op}}) e^{-\delta m t_s}
\ee
where $A$ and $B$ depend on the operator source-operator separation $t_{\rm{op}}$ 
of \eq{eq:ratio_def}.  
Note that we assume that only one excited state contributes in our fitting
range. For large enough $t_{\rm{op}}$, $A$ will become time independent
and the corresponding constant value provides the desired bare matrix element. 
In order to have stable fits we fix $t_{\rm{op}}$ and perform fits for
a range of $t_s \in [t_s^{min},t_s^{max}]$ and for a fixed value of
$\delta m$. Each choice of   
$\left\{ t_{\rm{op}},  [t_s^{min},t_s^{max}], \delta m \right\}$ 
will lead to a result for the extrapolated data. In order to estimate
the systematic error due to the excited
states contamination we thus vary all of them. For the preliminary
analysis presented here, 
four types of extrapolations denoted by 
I, II, III and IV, depending on the  choice of the parameters as
summarized in \tab{tab:fits}
\begin{table}
\begin{center}
    \begin{tabular}{| l | l | l | l |}
    \hline
    Type I & $t^1_{\rm{op}} \sim 0.66 \fm$ & $\sim[0.98,1.48 ] \fm$ 
    & $\delta m^1\approx 360~\mev$ \\
    Type II & $t^2_{\rm{op}} \sim 0.94 \fm$ & $\sim[ 1.15,1.48] \fm$ 
    & $\delta m^1\approx 360~\mev$ \\
    Type III & $t^1_{\rm{op}} \sim 0.66\fm$ & $\sim[ 0.98,1.48] \fm$ 
    & $\delta m^2\approx 600~\mev$ \\
   Type IV & $t^2_{\rm{op}}\sim 0.94\fm$ & $\sim[ 1.15,1.48] \fm$ 
    & $\delta m^2 \approx 600~\mev$ \\
    \hline
    \end{tabular}
\end{center}
\caption{Parameters used for the extrapolation to infinite source-sink
  separation. Two values of $\delta m$ are considered, one is fixed
  according to the   experimental mass spliting between $N$ and $N^\ast$, and the other
  is estimated from fits of the two-points function. In practice the
  values of $t_{\rm{op}},t_s^{min}$and $t_s^{max}$ are slightly varied
  on each ensemble in order to have stable fits.}\label{tab:fits}
\end{table}

\begin{figure}[h]
\begin{minipage}[ht]{7.5cm}
\vspace{+0.3cm}
\includegraphics[height=6cm,width=7.5cm]{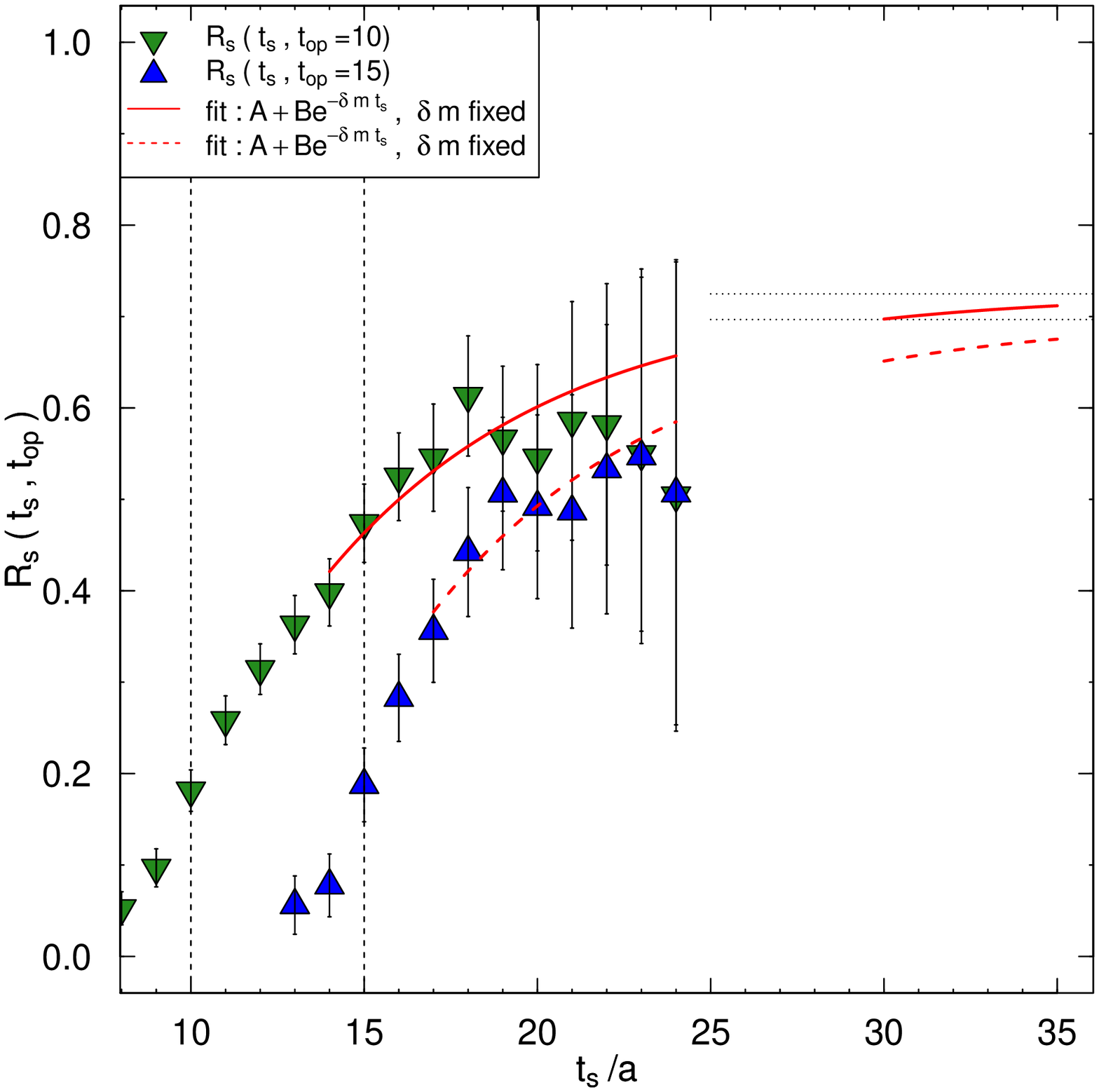}
\vspace*{-0.7cm}\caption{Extrapolation of 
$R_{O_s}(t_s,t_{\rm(op)})$ to large times $t_s$ for
two values of $t_{\rm{op}}$ and for $\delta m$ fixed to the
experimental mass splitting between the nucleon $N$ and 
the first excited state $N^\ast$. The gauge ensemble is
the same as the one used in Fig.~\protect\ref{fig:Rconn} }\label{fig:Rs_extrapolation}
\end{minipage}
\hspace{0.5cm}
\begin{minipage}[ht]{7.5cm}
\includegraphics[height=6cm,width=7.5cm]{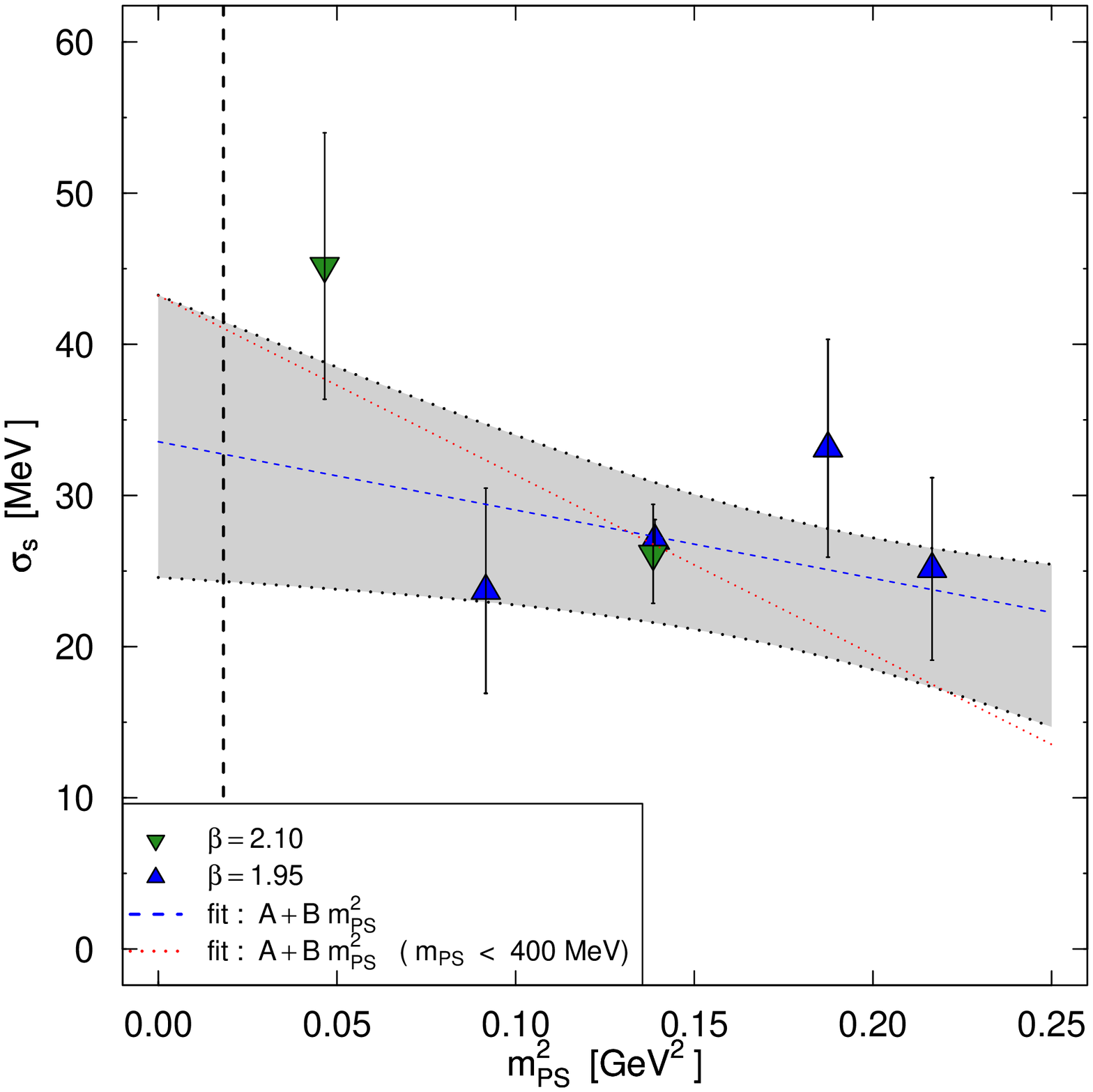}
\caption{Chiral behaviour of $\sigma_s$ as a function of
$\mps^2$. Data at two lattice spacings are shown as well as a linear
extrapolation (dotted blue line) and its corresponding error band. 
The physical point is denoted by a vertical
dashed line.}\label{fig:xfit_sigma_s}
\end{minipage}
\end{figure}

An example of such an extrapolation for  $t_{\rm{op}} = 0.82 \fm$ and
$t_{\rm{op}}=1.23\fm$ is shown in \fig{fig:Rs_extrapolation}. The
values of $A(t_{\rm{op}})$ are shown by horizontal black lines. We have repeated this procedure on every
ensemble resulting in the corresponding values of $\sigma_s$
for all four types of extrapolation considered, as can be seen
\tab{tab:fits}. Note that the values of $t_{\rm{op}}$ used in the table are smaller
than the one considered in  \fig{fig:Rs_extrapolation} in order to
have stable fits on all ensembles.
For the type I extrapolation we show in \fig{fig:xfit_sigma_s} the chiral behaviour of
$\sigma_s$ for all our ensembles employing two different lattice
spacings. As can be seen comparing the green ($a=0.064\fm$) and 
blue triangles ($a=0.082\fm$), 
lattice cut-off effects are not visible at the present level of accuracy.
The extrapolations to the physical pion
mass are carried through using two linear
fits in $\mps^2$ excluding or not data obtained for $\mps > 400~\mev$.
The spread of the results at the physical pion mass, 
using type I, II, III, IV allow us to estimate the systematic error coming from
excited states contamination. The difference between the two
chirally extrapolated values provides an estimate of the systematic uncertainty
associated with the chiral extrapolation. 
Our preliminary result reads $\sigma_s = 28(8)(10)~\mev$ where the first
error is statistical and the second is the systematic error stemming from
both excited states and chiral extrapolation. This value is fully compatible with the bound on $\sigma_s$ computed in ref.~\cite{Alexandrou:2013nda}.

In the case of the light $\sigma-$term, $\sigma_{\pi N}$, we have
performed the extrapolation in $t_s$ of type I to IV only for the disconnected part. 
For the
connected part, which is the dominant contribution, we do not have access
to the full source-sink time dependence of $R_{O_l}$. Instead, we 
consider two ways to estimate the systematic error due to the
excited states contamination in the connected part. 
In the case
denoted $C$ we use the results obtained with $t_s = 0.98 \fm$ and 
can not resolve  
the contribution of the excited states in the
connected part. In the case D, we used the only ensemble for which
data at several source-sink separation are available  ( $a=0.082\fm$ and 
$m_{\rm PS}\approx 355~\mev$ ) to estimate the excited states
contamination in the connected part by comparing results obtained at
$t_s= 0.98 \fm$ and $t_s= 1.47 \fm$. We deduce that the size of 
the excited states contribution is $18\%$. Assuming that this effect does not depend on 
the value of the pseudoscalar mass, we shift the lattice data on all the
other ensembles by $18\%$. We then perform chiral extrapolations 
using the results of cases $C$ and $D$. Combining theses results 
with the combinations methods I to IV for the disconnected part
allows us to at least estimate the systematic error due to the excited states contamination. 

In the case of $\sigma_{\pi N}$ the dependence as a function of $\mps^2$ is not negligible. 
In \fig{fig:xfit_sigma_l} we show our results for $\sigma_{\pi  N}$ as a function of 
$\mps^2$ and for two lattice spacings using type I extrapolations for
the disconnected parts and a fixed source-sink separation of $t_s=0.98
\fm$ for the connected part (previously refered as case C).
 We furthermore indicate the physical point by a
vertical dashed line. Using the well known baryon chiral perturbation
theory result $m_N = m_N^{0}  + c_0 \mps^2 + c_1 \mps^3$ with $c_1= -
\frac{3 g_A^2}{16 \pi f_{\pi}^2}$ and using the relation (true in
chiral limit) $\sigma_{\pi N} = \mps^2 \frac{\partial m_N}{\partial
  \mps^2} $, we show by a blue dotted line a fit of the data given by 
\be\label{eq:Op3_sigmal}
\sigma_l = \mps^2 ( c_0 + \frac{3}{2} c_1 \mps)\,
\ee
where the coefficient $c_0$ is the only fitting parameter and $c_1$ is fixed
and deduced from the chiral expansion in $\hbchipt$ of the nucleon
mass. 
Inspecting \fig{fig:xfit_sigma_l}, when
performing such a fit, even restricting ourselves to $\mps<400~\mev$, the fit does
not describe the data. In order to obtain a description of the 
data, we add a term $\mps^4$ to \eq{eq:Op3_sigmal} with a free coefficient 
that is to be fitted. This fit describes the data indeed 
satisfactory as shown in \fig{fig:xfit_sigma_l}. Also, in this case 
including or not data with pseudoscalar
meson masses larger than $400~\mev$ gives consistent results (see the red
dashed line and the  full black line in \fig{fig:xfit_sigma_l}).
We estimate the systematic error coming from the chiral extrapolation
by comparing the extrapolation of a linear  (obtained from \eq{eq:Op3_sigmal} setting
$c_1=0$) and quartic fit. We find 
$\sigma_{\pi N} = 37(2.6)(24.7)~\mev$ where the first error is 
statistical and the second is our estimate
of the systematic errors due to the chiral extrapolation and to the
excited states contamination.

\subsection{$y_N$ parameter}

\begin{figure}[h] 
\begin{minipage}[ht]{7.5cm}
\vspace{+0.0cm}
\includegraphics[height=6cm,width=7.5cm]{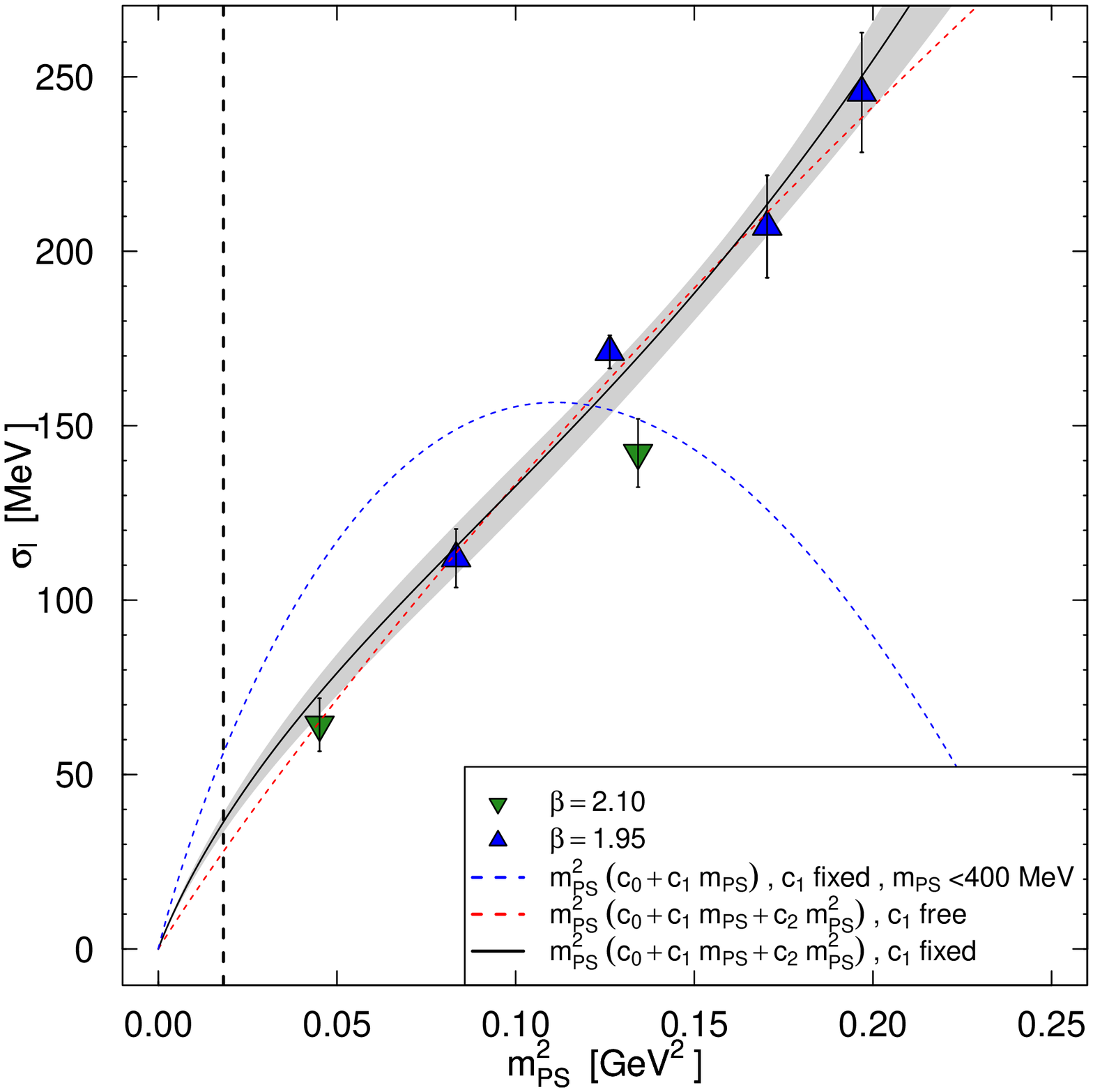}
\vspace*{-0.2cm}\caption{Chiral behaviour of $\sigma_s$ as a function of
$\mps^2$ after extrapolation of type I-C. We show data at two
lattice spacings. 
As explained in the text we show cubic  and quartic fits. The physical point is denoted by a
vertical dotted line.}\label{fig:xfit_sigma_l}
\end{minipage}
\hspace{0.5cm}
\begin{minipage}[ht]{7.5cm}
\includegraphics[height=6cm,width=7.5cm]{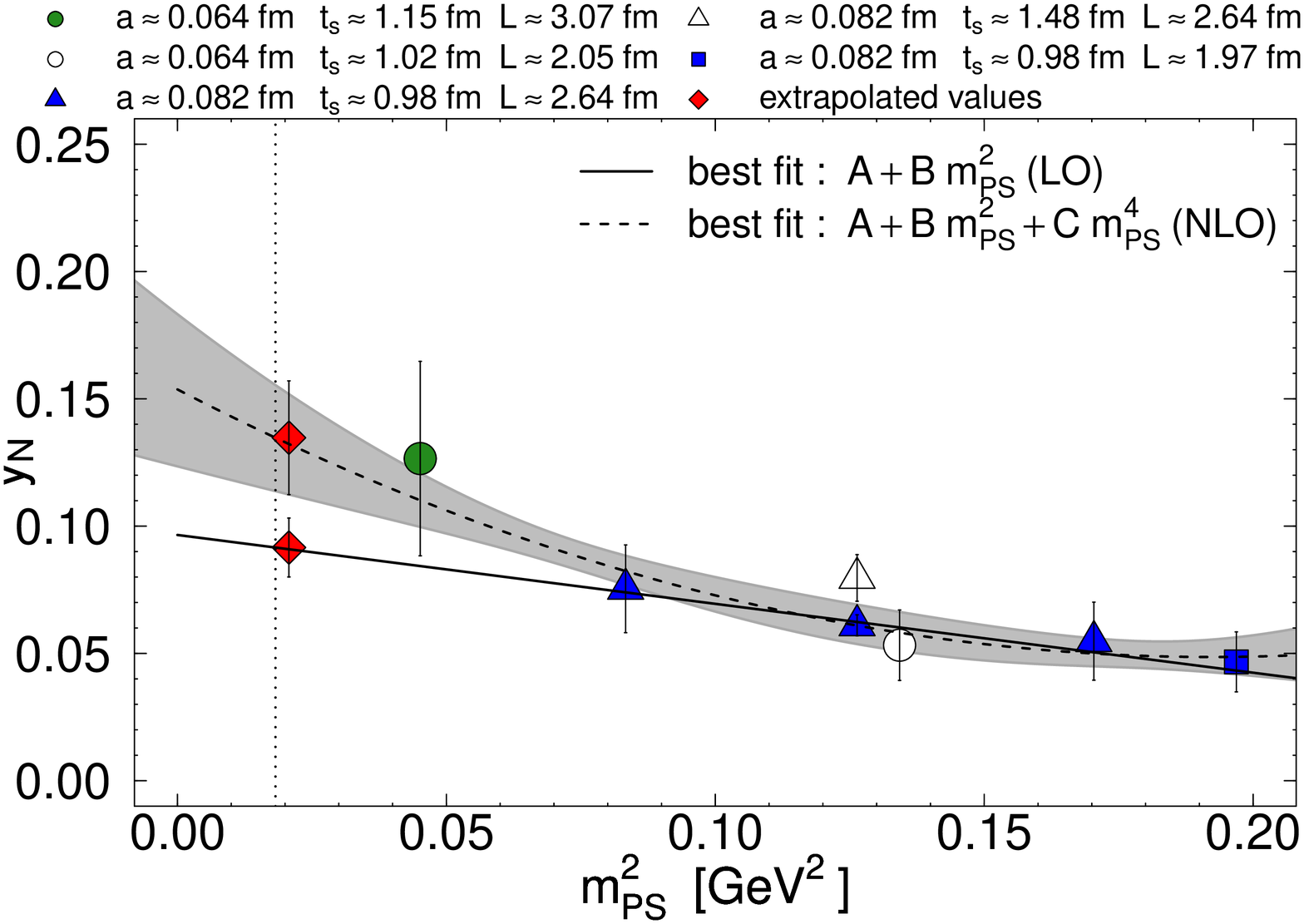}
\caption{Our results for $y_N$ as a function of 
$\mps^2$. for two values of the lattice spacing for several fixed
source-sink separation $t_{{\rm s}}$ as given in the legend. We extrapolate to the 
physical value of the pion mass using linear and 
quadratic fits in $\mps^2$. 
For the quadratic fit, we also show the corresponding error band.}\label{fig:xfit_yN}
\end{minipage}
\end{figure}

We summarize here briefly our analysis of the $y_N$ parameter. Contrary
to the $\sigma-$terms, $y_N$ can be obtained directly by a ratio
of three point function. We show in \fig{fig:xfit_yN} our results for $y_N$ as a
function of $\mps^2$. Data for several source-sink separation are
shown as well as for two lattice spacings. We observe that data obtained at different
source-sink separations (filled and empty triangle) indicate that the
excited states contamination is non negligible and about $32\%$. The
excited states contamination is thus small compare to the excited
states contamination in the determination of $\sigma_s$ which must
partly cancel in the ratio used to compute $y_N$.
We also show the result
obtained for a different lattice spacing in the same range of
pseudoscalar mass (empty circle) to indicate that lattice cut-off effects
are not a dominant source of systematic errors. We then perform a leading
order (linear in $\mps^2$) and an extrapolation 
adding a $\mps^4$ term to the fit formula to estimate the error
introduced by the chiral extrapolation. Note that the empty symbols
are not included in the fits and are only shown to estimate systematic
errors. Our final result is (including systematic errors) 
$y_N=0.135(46)$. We refer to \cite{Alexandrou:2013nda} for
more details on this analysis.

\section{Conclusion}

In this proceedings contribution we have presented our results
concerning the nucleon light and
strange $\sigma-$terms. We discussed in  detail the  analysis of the systematic
effects in a direct computation of these matrix elements. In particular, 
we showed that a large excited states
contamination is present in the determination of the scalar quark matrix
elements of the nucleon. 
This is the first time that a careful assessment of excited states
contributions to the $\sigma-terms$  has been carried out resulting into
a systematic which is larger than the statistical error and must thus
be considered in the evaluation of these quantities.
We have also shown that lattice cut-off effects do
not affect our results at the present level of accuracy. 
Another source of systematic effects 
in the case of $\sigma_{\pi N}$, which we have taken into 
account, is the chiral
extrapolation.
Our results for the $\sigma$-terms read $\sigma_{\pi N} = 37(2.6)(24.7) ~\mev$ and
$\sigma_s=28(8)(10)~\mev$. 
We summarised our recent analysis of the $y_N$ parameter and found
$y_N=0.135(46)$ including systematic
errors. This estimate provides a reliable input for dark matter models and experimental  searches.

\vspace*{-0.5cm}
\section*{Acknowledgments}
\vspace*{-0.5cm}
This work was performed using HPC resources provided by the JSC 
Forschungszentrum J\"ulich on the Juqueen supercomputer. It is supported in 
part by  the DFG Sonder\-for\-schungs\-be\-reich/ Trans\-regio SFB/TR9. 
Computational resources where partially provided by the Cy-Tera 
Project NEA Y$\Pi$O$\Delta$OMH/$\Sigma$TPATH/0308/31 funded
by the Cyprus Research Promotion Foundation (RPF).  A. V is supported
by the Cyprus RPF under the grant $\Pi$PO$\Sigma$E$\Lambda$KY$\Sigma$H/NEO$\Sigma$/0609/16.


\vspace*{-0.5cm}
\begin{thebibliography}{99}

\bibitem{Alexandrou:2013nda} 
  C.~Alexandrou, M.~Constantinou, S.~Dinter, V.~Drach, K.~Hadjiyiannakou, K.~Jansen, G.~Koutsou and A.~Vaquero,
  arXiv:1309.7768 [hep-lat].


\bibitem{Dinter:2012tt} 
  S.~Dinter, V.~Drach, R.~Frezzotti, G.~Herdoiza, K.~Jansen and G.~Rossi,
  JHEP {\bf 1208}, 037 (2012)
  [arXiv:1202.1480 [hep-lat]].


\bibitem{Jansen:2008wv} 
  K.~Jansen {\it et al.}  [ETM Collaboration],
  Eur.\ Phys.\ J.\ C {\bf 58}, 261 (2008)
  [arXiv:0804.3871 [hep-lat]].

\bibitem{Alexandrou:2012py} 
  C.~Alexandrou, V.~Drach, K.~Hadjiyiannakou, K.~Jansen, G.~Koutsou, A.~Strelchenko and A.~Vaquero,
  arXiv:1211.0126 [hep-lat].

\bibitem{Engelhardt:2012gd} 
  M.~Engelhardt,
  arXiv:1210.0025 [hep-lat].
\bibitem{Oksuzian:2012rzb} 
  H. Ohki,{ \it et al.}  [JLQCD Collaboration],
  arXiv:1208.4185 [hep-lat].
\bibitem{Bali:2012qs} 
  G.~S.~Bali, P.~C.~Bruns, S.~Collins, M.~Deka, B.~Glasle, M.~Gockeler, L.~Greil and T.~R.~Hemmert {\it et al.},
*  
  Nucl.\ Phys.\ B {\bf 866}, 1 (2013)
  [arXiv:1206.7034 [hep-lat]].
\bibitem{Alexandrou:2013wca} 
  C.~Alexandrou, M.~Constantinou, V.~Drach, K.~Hadjiyiannakou, K.~Jansen, G.~Koutsou, A.~Strelchenko and A.~Vaquero,
  arXiv:1309.2256 [hep-lat].


\bibitem{Baron:2010bv}
  R.~Baron, P.~.Boucaud, J.~Carbonell, A.~Deuzeman, V.~Drach, F.~Farchioni, V.~Gimenez, G.~Herdoiza {\it et al.},
  JHEP {\bf 1006}, 111 (2010).
  [arXiv:1004.5284 [hep-lat]].

\bibitem{Alexandrou:2012gz} 
  C.~Alexandrou, M.~Constantinou, S.~Dinter, V.~Drach, K.~Hadjiyiannakou, K.~Jansen, G.~Koutsou and A.~Strelchenko {\it et al.},
  PoS LATTICE {\bf 2012}, 163 (2012)
  [arXiv:1211.4447 [hep-lat]].


\end{thebibliography}
\end{document}